\documentclass[sigconf,screen]{acmart}
\usepackage{booktabs} % For formal tables

\AtBeginDocument{%
  }

\graphicspath{{figures/}{pictures/}{images/}{./}} % where to search for the images

\usepackage{amsmath}
\usepackage{algorithm}

\usepackage{color}
\usepackage{bbding}
\usepackage{booktabs}

\usepackage{enumitem}
\setlist{noitemsep,parsep=0pt,partopsep=0pt, leftmargin=10pt} 
\usepackage{listings}
\usepackage{multirow}
\usepackage{colortbl}
\usepackage{graphicx}
\usepackage{mwe} 
\usepackage{fontawesome}
\usepackage{duckuments}

\AtBeginDocument{

}
\newcommand{\eg}{{\it e.g.,\ }}

\newcommand{\ie}{{\it i.e.,\ }}

\newcommand{\media}{Live Artifacts}
\newcommand{\system}{LiveCanvas}

\definecolor{Restrict}{HTML}{eee0da}
\definecolor{Expand}{HTML}{d3e5ef}
\definecolor{Refine}{HTML}{dbeddb}
\definecolor{Organize}{HTML}{e8deee}
\definecolor{Follow-up}{HTML}{fdecc8}
\definecolor{Initial}{HTML}{ffe2dd}
\definecolor{Human}{HTML}{fadec9}
\definecolor{AI}{HTML}{d3e5ef}
\definecolor{Direct}{HTML}{eee0da}
\definecolor{Indirect}{HTML}{dbeddb}
\definecolor{Visual Brush Selection}{HTML}{f5e0e9}
\definecolor{Widget Selection}{HTML}{fdecc8}
\definecolor{Inline Annotation}{HTML}{dbeddb}
\definecolor{Interactive Visual Click}{HTML}{ffe2dd}
\definecolor{Flowchart Manipulation}{HTML}{e8deee}
\definecolor{Tree Manipulation}{HTML}{fadec9}
\definecolor{Text Brush Selection}{HTML}{d3e5ef}
\definecolor{Text Input or Editing}{HTML}{ffffcc}
\definecolor{Inline Highlighting}{HTML}{e9fbe0}
\definecolor{Viewpoint Navigation}{HTML}{e8deee}
\definecolor{Sketch}{HTML}{c8e3f3}
\definecolor{Spreadsheet Manipulation}{HTML}{f1f0ef}
\definecolor{Speech}{HTML}{f1f0ef}
\definecolor{Global Software Manipulation}{HTML}{f1f0ef}
\definecolor{Drag and Drop}{HTML}{f1f0ef}

\definecolor{tablerowcolor}{rgb}{0.667,0.667,0.667 }
\definecolor{tablerowcolor2}{rgb}{0,0,0}
\definecolor{visual}{HTML}{e8efd9}
\definecolor{motion}{HTML}{fde7d5}
\definecolor{narrative}{HTML}{e2dce9}
\definecolor{audio}{HTML}{d6ebf2}
\definecolor{bluecrayola}{rgb}{0.12,0.46,1.0}

\begin{document}
\title[Live Artifacts: Authoring Dynamic Media via Live Layers Encapsulating Generative Specifications]{Live Artifacts: Authoring Dynamic Media via Live Layers Encapsulating Generative Specifications}

\author{Leixian Shen}
\authornote{Work done during internship at Microsoft Research.}
\orcid{0000-0003-1084-4912}
\affiliation{%
  \institution{Microsoft Research}
  \city{Redmond}
  \state{WA}
  \country{USA}
}
\affiliation{%
  \institution{The Hong Kong University of Science and Technology}
  \city{Hong Kong SAR}
  \country{China}
}
\email{lshenaj@connect.ust.hk}

\author{Haotian Li}
\orcid{0000-0001-9547-3449}
\affiliation{%
  \institution{Microsoft Research Asia}
  \city{Beijing}
  \country{China}
}
\email{haotian.li@microsoft.com}

\author{Hugo Romat}
\orcid{0000-0002-9194-8889}
\affiliation{%
  \institution{Microsoft Research}
  \city{Redmond}
  \state{WA}
  \country{USA}
}
\email{romathugo@microsoft.com}

\author{Fanny Chevalier}
\orcid{0000-0002-5585-7971}
\affiliation{%
  \institution{Microsoft Research}
  \city{Redmond}
  \state{WA}
  \country{USA}
}
\affiliation{%
  \institution{University of Toronto}
  \city{Toronto}
  \state{ON}
  \country{Canada}
}
\email{fanny@dgp.toronto.edu}

\author{Nicolai Marquardt}
\orcid{0000-0002-5473-2448}
\affiliation{%
  \institution{Microsoft Research}
  \city{Redmond}
  \state{WA}
  \country{USA}
}
\email{nicmarquardt@microsoft.com}

\author{Nathalie Riche}
\orcid{0000-0003-1759-7512}
\affiliation{%
  \institution{Microsoft Research}
  \city{Redmond}
  \state{WA}
  \country{USA}
}
\email{nath@microsoft.com}

\begin{abstract}

We frame \textit{Live Artifacts} as a class of persistent generative media between static assets and interactive software.
Unlike conventional generative outputs that collapse into static files, \media~ retain their generative logic as a persistent media property, enabling continuous context-dependent regeneration. Time, location, or live data become part of their generative specifications, initiating coordinated updates across modalities (\eg adapting text, visuals, and audio together) while preserving composition, semantics, identity, and cross-modal coherence. To facilitate experimentation with this medium, we present \system, an authoring system that reconceptualizes visual layers as live generative specifications with explicit execution modes and constrained dependencies. Creators orchestrate dynamic behaviors and manage generative persistence within a visual canvas rather than through programming, defining what remains stable, what can change, and how changes propagate. We evaluate \media~ through a gallery of responsive examples and a qualitative study with six professionals, finding that \system~facilitates a shift from composing static outputs to crafting persistent generative artifacts while remaining aligned with familiar authoring practices.

\end{abstract}

%\begin{CCSXML}
%<ccs2012>
%   <concept>
%       <concept_id>10003120.10003121.10003129</concept_id>
%       <concept_desc>Human-centered computing~Interactive systems and tools</concept_desc>
%       <concept_significance>500</concept_significance>
%       </concept>
%   <concept>
%       <concept_id>10003120.10003121.10003124</concept_id>
%       <concept_desc>Human-centered computing~Interaction %paradigms</concept_desc>
%       <concept_significance>500</concept_significance>
%       </concept>
% </ccs2012>
%\end{CCSXML}



 \begin{teaserfigure}
 \centering
     \includegraphics[width=\linewidth]{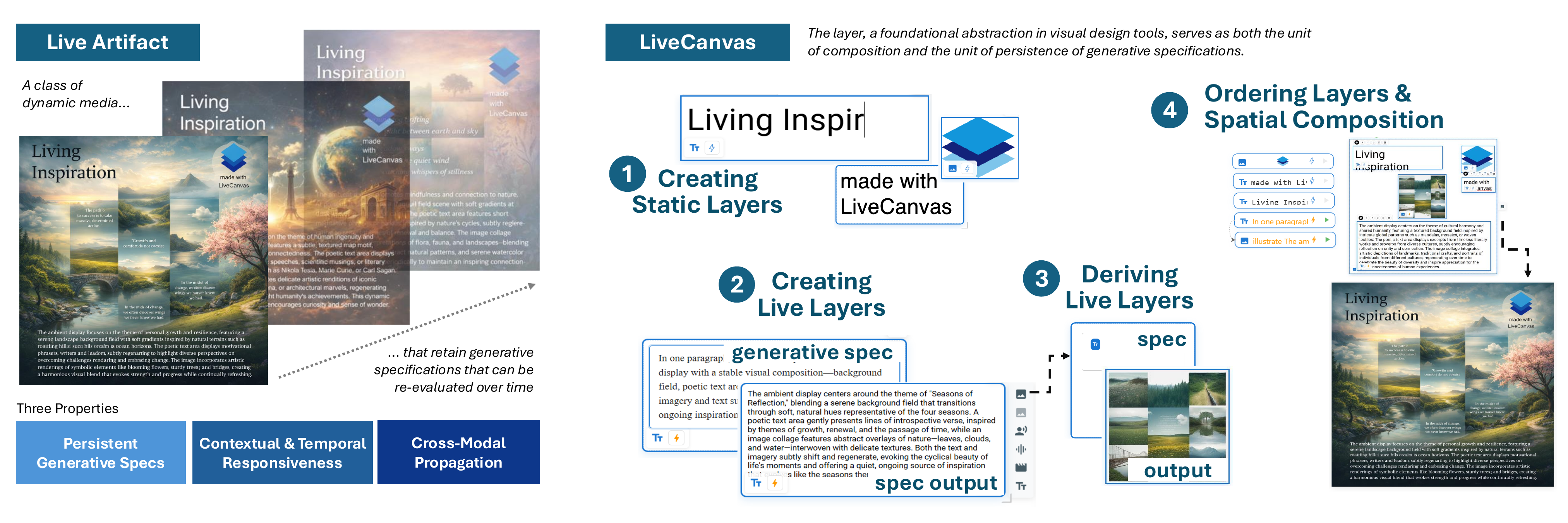}
      % \vspace{5px}
     \caption{\textit{\media~} encapsulate persistent generative specifications as an intrinsic property of the media itself, driving continuous re-evaluation and cross-modal propagation. \textit{\system~} operationalizes this concept as a media-first authoring interface. By elevating familiar visual layers to encapsulate generative specifications, creators can visually orchestrate static and live media, leveraging cross-modality propagation between layers.}
     \vspace{10px}
 \label{fig:teaser}
 \end{teaserfigure}

\maketitle

\section{Introduction}
\label{sec:intro}

The rapid advancement of Generative AI (GenAI) is lowering the barrier to multimedia creation. Today, people can produce high-quality images, text, audio, and video by expressing intent in natural language, often without needing to learn professional design tools or write software~\cite{Subramonyam2023, IAI, Riche2025}. 
Yet, current creation workflows typically treat GenAI as an ephemeral, one-shot rendering engine. Once media is generated, the underlying generative logic is discarded, leaving behind static outputs that cannot adapt or respond to changing context.

At the opposite end of the spectrum lies code. Software can encode persistent logic, reactivity, and states, enabling potentially complex systems that evolve over time and respond to user input or external data. Recent LLM-assisted programming environments (\eg Cursor~\cite{cursor}, Github Copilot~\cite{github_copilot}) lower the barrier to entry by using natural language to generate and modify software. However, authoring such systems requires an engineering mindset oriented around control flow, state management, debugging, and deployment. For many creators, this represents not just a technical hurdle but a shift in how problems are framed and solved, not necessarily aligned with the familiar ways designers typically think about composing and refining media~\cite{Satyanarayan2019a, Liu2023e}.

This contrast reveals a large and underexplored gap between static media and full‑fledged software. Advances in GenAI invite reconsideration of what media artifacts could be if persistence, cross‑modal coherence, and contextual/temporal responsiveness can be treated as intrinsic properties of media itself. 
For example, a promotional display may preserve a fixed layout and product identity while regenerating imagery based on viewer location and interests, enabling personalization at scale. 
Or a short‑form social media post may continuously regenerate visuals, captions, and audio around a recognizable meme to retain attention. 
Such artifacts require generative logic that persists across sessions, responds to external context, and where changes propagate across modalities. We refer to this emerging class of dynamic media as \textit{Live Artifacts}.

Authoring \media~ introduces a central tension: managing the boundary between what should remain stable (\eg a brand logo, a character's identity, a layout) and what should evolve continuously --- and how. Historically, resolving this required abandoning the design canvas for software development. 
Our key insight is that recent advances in generative AI obviate the need for this transition. 
Rather than shifting to a visual programming or code generation mindset, we propose to retain a media‑first abstraction by extending the capabilities of visual layers to support generative logic.
Visual layers, the foundational abstraction of media creation tools, already encode how authors reason about structure through spatial composition, hierarchy, grouping, and dependencies~\cite{Masson2025, Siddiqui2025, Ma2022}. This existing logic makes layers an ideal foundation for \media~authoring: layers align with designers' mental model while providing a structured backend that can be augmented for generative re-evaluation.

To facilitate experimentation with \media, we present \system, an authoring system building upon the concept of layers, augmenting them with the ability to encapsulate \textit{generative specifications} directly within the visual canvas, turning them into ``live layers''. Authors decide which layers are static, and which are live, \ie re-evaluated at each consumption. Authors can establish dependencies between layers through which changes organically propagate across the artifact, including updates driven by external context or time.  These links are intentionally constrained: they describe how content flows between multimodal layers, without introducing loops, hidden states, or complex control flow. 
Unlike node-based visual programming interfaces (\eg ComfyUI~\cite{comfyui}) that separate pipeline logic from the final output, our linked-layer approach spatially grounds all generative logic directly within the artifact itself. As a result, authors work by arranging, linking, and adjusting visible elements on a canvas, rather than reasoning about programs and execution order.

To illustrate the expressive range enabled by this approach, we present a gallery of \media~ spanning diverse scenarios like information displays, narrative media, and ambient experiences. These examples demonstrate how persistent generative relationships allow media to evolve over time while maintaining coherence across modalities and contexts. To understand how creators engage with this approach, we conducted a qualitative study with six multimedia professionals using \system. Rather than measuring task efficiency, the study examines how authors reason about persistence, dependencies, and change when working with live layers, and how their mental models shift from composing static outputs to orchestrating generative behaviors within a visual canvas.

In summary, this paper makes the following contributions:

\begin{itemize}
    \item \media, a class of dynamic media artifacts in which re-evaluation logic persists in generative specifications, propagates across modalities, and responds to context, occupying a space between static media and software.
    \item \system, a system for authoring generative specifications within a media‑first, visual workflow by making multimodal layers live and linking them through constrained dependencies.
    \item A gallery of examples and findings from a preliminary qualitative study showing how creators author and reason about the live, generative media using \system.
\end{itemize}

Together, these contributions outline a design space for live, generative media and present a visual authoring system that brings capabilities traditionally associated with software into media creation workflows.

\section{Related Work}

We first review research on dynamic and interactive media to situate Live Artifacts in this space; we then review content creation tools leveraging generative AI and delve into authoring paradigms leveraging layers as a key abstraction.

\subsection{Dynamic and Interactive Media}
Computational notebook environments like Jupyter Notebooks~\cite{Jupyter} and Observable~\cite{Observable} are forms of interactive documents in which analysts, data scientists and programmers modify and re-execute code alongside prose. These environments can be further augmented with bidirectional code--visualization bindings~\cite{Wu2020B2}. 
Victor~\cite{Victor2011} argued more broadly for explorable explanations, which provides the ability for readers to manipulate parameters and observe consequences in rendered figures and text. This vision was realized in interactive multiverse analyses~\cite{dragicevic2019emars}, interactive ML explanations~\cite{Distill}, and markup-based languages producing interactive media~\cite{Conlen2021}.

Recent toolkits push this even further: Living Papers~\cite{Heer2023} compiles articles into reactive web documents with executable code, and CrossData~\cite{10.1145/3491102.3517485} maintains persistent text-to-data links where changes propagate automatically. 
Targeted at end-users, frameworks such as Mavo~\cite{Verou2016}
and Varv~\cite{Borowski2022} let non-programmers create reactive applications through declarative data structures and defining higher level concepts.
Towards distributed platforms, Streamlit~\cite{Streamlit} facilitates reactive data sharing across web apps, and Webstrates~\cite{Klokmose2015} showed that documents can serve as live computational substrate by targeting the underlying DOM-level state. 
In parallel, the visualization research community has studied mechanisms for interactive data‑driven storytelling~\cite{Boy2015, Segel2010, Stopler2016, McKenna2017}. This also motivates work for narrative sequencing~\cite{Satyanarayan2014}, scrollytelling~\cite{Satyanarayan2014b,sultanum2021leveraging} (unfolding content as users scroll through a webpage), and data videos~\cite{dataplaywright, DVSurvey}.

Across these approaches, media-centric systems offer powerful support for spatial composition and direct manipulation, while logic-centric systems contribute persistent specifications. Live Artifacts build on and expand on those, embedding generative specifications directly within spatial layers, preserving the underlying generative specification as part of the artifact itself. 
Different from manual regeneration or deterministic models, this is a complementary class of media allowing re-evaluation triggered by context changes or timed events.

\vspace{-0.2cm}
\subsection{Authoring Dynamic Media with GenAI}

Authoring dynamic media with generative AI requires balancing two concerns: arranging content spatially and managing generative behavior that can change the content over time. Most existing tools support one of these well, but not both together.

Media‑centric tools such as Photoshop~\cite{Photoshop} and Figma~\cite{Figma} emphasize visual composition and direct manipulation. Even when augmented with generative AI features (\eg Generative Fill~\cite{Generative_Fill}), generation is typically treated as a one‑time operation whose results are flattened into static pixels.
Yet, constraint-based predecessors --- such as the seminal Sketchpad~\cite{Sutherland1963} or ThingLab~\cite{Borning1981} --- showed that persistent declarative relationships enable powerful propagation within documents.

Logic-centric systems foreground generative process over spatial composition. Node‑based environments such as PromptChainer~\cite{Wu2022PromptChainer}, ComfyUI~\cite{comfyui}, and  Node‑RED~\cite{NodeRED} let authors define declarative pipelines. Structured creative tools such as Opal~\cite{Liu2022Opal}, Luminate~\cite{Suh2024Luminate}, and Dreamsheets~\cite{Almeda2024DreamSheets} support systematic exploration of generative design spaces. And Canvas-based systems like WorldSmith~\cite{Dang2023WorldSmith}, ImaginationVellum~\cite{Marquardt2025}, and Spellburst~\cite{Angert2023} further blend persistent spatial layout with iterative generation pipelines. 
However, these tools typically separate the generative pipeline from the media it produces, possibly requiring interpretive overlays, provenance graphs, or temporal replay~\cite{Jiang2023, Suh2023a, Marquardt2025} to interpret and relate process to output in post-hoc explorative sensemaking.
At the other end of the spectrum, 
Code‑centric approaches (\eg Cursor~\cite{cursor}, GitHub Copilot~\cite{github_copilot}) offer broad computational expressiveness for building dynamic media but move authoring away from direct manipulation of visible objects with immediate feedback~\cite{Shneiderman1997}.

Across these approaches, it remains difficult to both compose media visually and retain generative behavior as a first‑class part of the artifact. \media{} address this gap: generative specifications are embedded directly within spatial layers on a visual canvas, affording both direct manipulation and re-evaluation.

\vspace{-0.2cm}
\subsection{Layers for Structuring Media Creation}
Layering is the dominant organizational metaphor in digital media creation, popularized by tools like Photoshop and Figma.
Rooted in early image compositing~\cite{porter1984compositing, smith2001digital}, the layer model structures media as a spatially ordered stack, where each layer is rendered individually before being merged.
The core affordance is non-destructive, separable editing: elements can be isolated, modified, and blended without altering underlying content.

Valued for their qualities as a cognitive aid for organizing elements and a computational tool for isolated rendering of visual effects, layers have been investigated by researchers across a range of media and interaction contexts, including the authoring of stylized 3D animation~\cite{Ma2022}, immersive 3D painting~\cite{Yu2024}, and even text~\cite{Masson2025,Siddiqui2025}. Layering has also been applied to generative pipelines; for instance, LayerAnimate~\cite{Yanga} introduces a layer‑aware diffusion framework that allows creators to exercise fine‑grained, layer‑level control over video generation.

Prior work demonstrates the power of layers as a conceptual bridge between the creator's mental models and the underlying rendering pipelines. Layers provide a natural means of isolating and encapsulating distinct media aspects or properties, making them an ideal technique for authoring emerging media such as Live Artifacts.  Our work examines how layers might be augmented to encapsulate generative specifications within dynamic media artifacts, thereby enabling their continuous re-evaluation.

\section{\textit{Live Artifacts}}

We use the term \textbf{\media}~ to describe a class of media artifacts composed of both static media and generated media components, where the generative specifications governing these components --- and the relationships between them --- are retained as inherent properties of the artifact. These specifications can be re‑evaluated over time, allowing the artifact to produce different outputs while preserving an underlying skeleton. Live artifacts are best seen as a form of dynamic media than programs, as they do not support open-ended interaction logic or control flow.

\subsection{Core Properties}

We identify three core properties (C1-C3) for \media:

\paragraph{C1: Persistent generative specifications.}
Unlike one‑shot generative outputs, \media~ preserve their generative specifications as first‑class materials. The artifact is therefore not only a set of pixels or text, but also encapsulate generative specifications to be re‑evaluated over time. Authors determine which elements remain static and which remain live. This enables artifacts to preserve selected characteristics (e.g. a recognizable layout incorporating static media such as a logo or title), while incorporating new content when specifications are re-evaluated.

\paragraph{C2: Contextual and temporal responsiveness.}
Live Artifacts can respond to external data such as time, location, or any web data sources by incorporating URLs directly in the generative specifications of its components. This enable personalization at scale, such as adapting visuals to the locale of the viewer. Temporal responsiveness refers to the evolution of the artifact over time (by automatically triggering re-evaluation over a time period), or at each consumption. 

\paragraph{C3: Cross‑modal propagation.}
\media~ explicitly encode relationships between layers by enabling authors to refer to a parent layer in the specification of the child layer. Changes to one parent layer (\eg text) propagate to others (\eg imagery or audio), enabling the artifact stay coherent across modalities as it evolves. These relationships are intentionally limited (\eg each layer refers to a single parent) to avoid complex or cyclic propagation flows.

\begin{figure}[t]
  \centering
    \includegraphics[width=\linewidth]{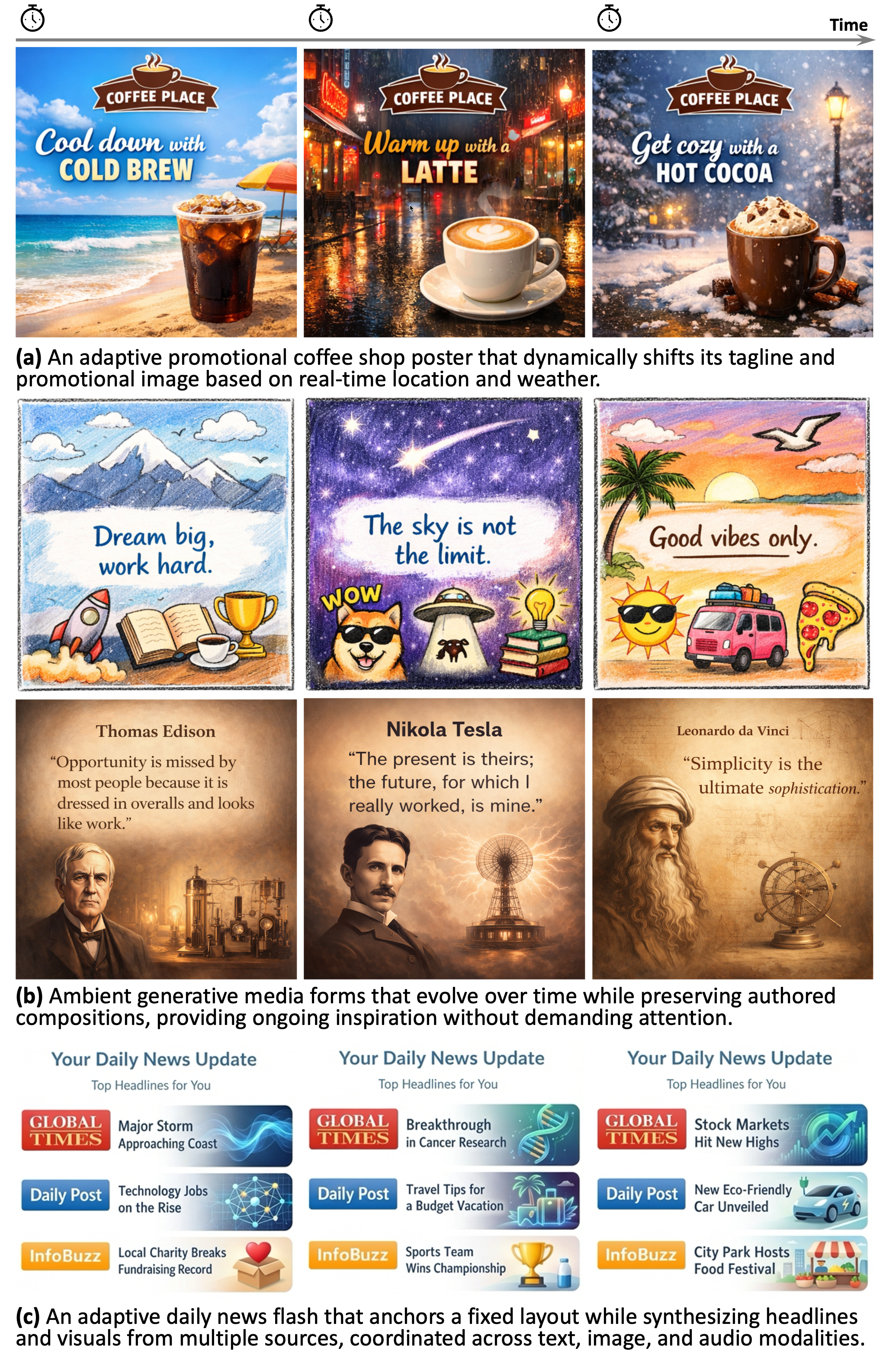}
    \vspace{-23px}
    \caption{Examples of Live Artifacts.}
\label{fig: example1}
   \vspace{-13px}
\end{figure}

\subsection{Examples} 

We illustrate \media~ through examples (\autoref{fig: example1}) and clarify how they differ from both static media and traditional software.

\textbf{Context-aware promotional content.}
Consider a digital promotional display advertising a coffee beverage. 
For promotional media, attracting and sustaining viewer attention is critical, and personalization at scale can significantly increase relevance. For beverages, this may mean reflecting where and how viewers are likely to want one. 
A Live Artifact can contain a tagline and promotional image regenerating based on viewer's location and weather, yet fix the composition to include the coffee shop logo and maintain a visual identity (\autoref{fig: example1}-a). 
Location and weather are referenced directly within the generative specification that produces the tagline, then propagated to the specification generating the image, allowing this contextual information to propagate across modalities. The images are then generated by an AI model: even when viewed under similar conditions, the produced artifact differs while still preserving key characteristics and layout, ensuring repeated encounters feel fresh and continue to attract attention.

Although this adaptation may appear as a minor extension of static media, achieving it in practice would either require extensive manual duplication: authoring and managing separate background assets for different locations and conditions; or turning to application development. In the latter case, this typically means writing or generating code to retrieve context, orchestrate calls to image generation models, manage variants, and integrate the results back into a composed layout. Even with advanced ``vibe‑coding'' tools or developer assistance, this introduces non‑trivial effort and iteration. In a Live Artifact, the same outcome is achieved by re‑evaluating the artifact using contextual references (\eg location and weather) and propagating that result through dependencies.

\textbf{Ambient generative media for inspiration.}
Consider an ambient Live Artifact on a wall display. Ambient media often aims to provide ongoing inspiration without demanding attention, balancing novelty with visual stability over extended periods of time. The artifact has a specific composition that remains stable over time: a large background field, a region for a short poetic line of text, and a small collage of images with a distinct look and feel. This composition defines the artifact’s identity (\autoref{fig:teaser} and \autoref{fig: example1}-b). What changes is the  content that inhabits these regions. At regular intervals, the content is re-evaluated given the generative specifications encapsulated in each live layer. 
As a Live Artifact, the composition is authored once. Generative specification and its logic of propagation is preserved, enabling regeneration within the user specified bounds.

\textbf{Personalized audio news update.}
As a third example, consider a custom daily news flash, synthesizing top headlines of three news media outlets in a fixed layout (\autoref{fig: example1}-c). The specification defines whether using global or local news, and an audio layer summarizing the updates makes it easier to listen to the news headlines on the go.  Live Artifacts facilitate synthesizing those multiple news sources and orchestrating the creation of new content (visual, audio) all without the need of any programming. 

These examples show how Live Artifacts combine persistent generative logic, bounded cross‑modal propagation, and selective contextual or temporal responsiveness to enable change while remaining recognizably the same media artifact.

\subsection{Live Artifact vs. Template and Application}

While Live Artifacts share structural similarities with dynamic media templates (\eg parameterized Figma components~\cite{figmaVariants} or After Effects MOGRTs~\cite{adobeMOGRTs}), they differ fundamentally in their underlying logic and binding mechanisms. 
Traditional templates rely on deterministic slots and pre-authored assets: the same input always produces the exact same output, and cross-modal propagation (\eg matching text styling to an inserted image) requires explicit, hard-coded scripting. In contrast, Live Artifacts replace rigid placeholders with persistent generative specifications. Rather than swapping assets within a static container, a Live Artifact uses persistent generative logic to synthesize new content and relies on cross-modal propagation --- rather than imperative code --- to propagate changes (such as changes in ``weather'') across text, audio, and visuals. By embracing the inherent non‑determinism of generative models, Live Artifacts trade predictability for expressive range, with outcomes that emerge rather than being fully specified in advance.

Live Artifacts overlap with software applications in that both retain authored logic across sessions, respond to external data, and evolve over time. 
A Live Artifact may regenerate itself based on time or external data rather than being regenerated from scratch, moving beyond the finality of static media and adopting some dynamic properties traditionally associated with software. 
Despite this overlap, Live Artifacts differ from programs in purpose and abstraction. 
Programs are organized around functionality and general behavior, relying on programming constructs (\eg state, events, and control flow) to ensure correct execution. This expressiveness enables unbounded flexible behavior, but it requires authors to reason explicitly about execution, state, and failure cases, shifting their focus from envisioning a media outcome to implementing programmatic logic and engineering workflows.

\begin{figure*}[t]
  \centering
    \includegraphics[width=\linewidth]{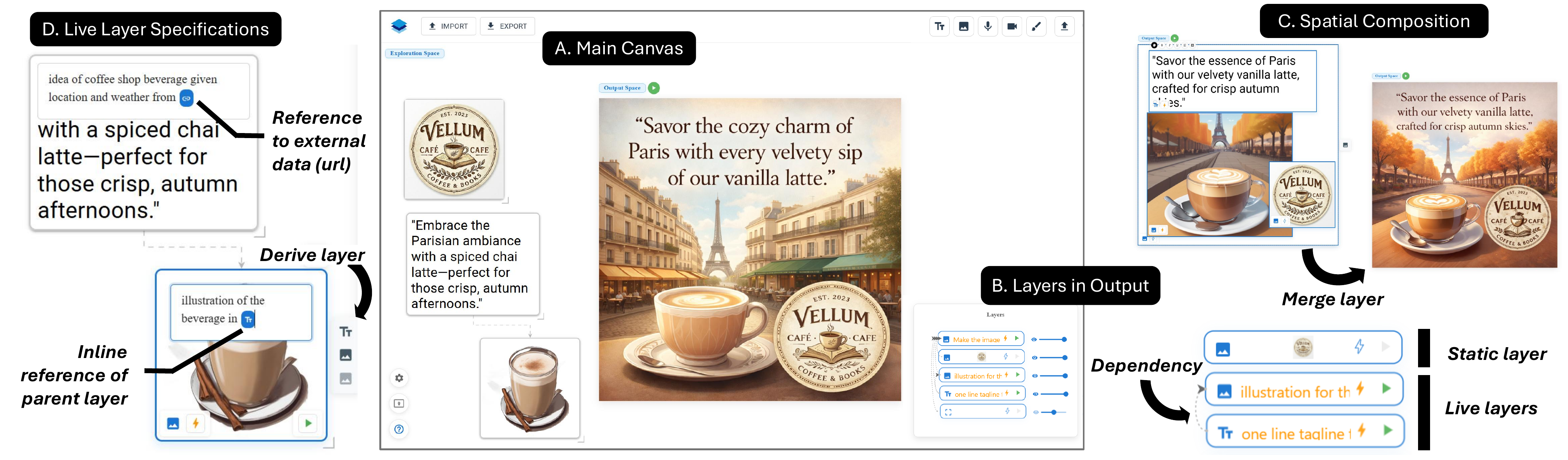}
     % \vspace{-5px}
    \caption{
    \system~Interface. 
    (A) The main canvas contains layers used for Exploration Space or placed in the Output Space (the live Artifact);
    (B) The layer stack lists layers in Output Space, including static layers (\eg logo) and live layers (\eg tagline, illustration), with arrows noting dependencies for propagation;
    (C) Layers are spatially arranged in Output Space, and a Merge Layer can provide a unified visual output;
    (D) Live layers encapsulate generative specifications that may include references to external data for contextual responsiveness; and inline references of the parent layer for cross-modal propagation.}
\label{fig: interface}
  % \vspace{-5px}
\end{figure*}

Live Artifacts are a form of media. They include static and dynamic media components, preserving the generative specifications of dynamic media as properties. Dependencies describe how content is derived or propagated between these components, and temporal change concerns when content is re-evaluated, rather than how general processes unfold through control flow. Live Artifacts can be understood and inspected through its visible structure directly rather than as an output of an execution. Live Artifacts are not intended to support complex interaction logic, long‑lived internal state, or procedural workflows. Extending them in this direction would shift the model away from media, toward general‑purpose computation as it would require reintroducing abstractions characteristic of programs.

\section{Authoring Live Artifacts with \system}

Having defined \media{} and their core properties, we now describe
LiveCanvas, a \textbf{media-first authoring interface} that enables non-programmers to create Live Artifacts without shifting to an engineering mindset. 
Layers serve as both the unit of composition and the unit of persistence of generative specifications in LiveCanvas (\autoref{fig: interface}). 
Some layers are authored to remain static (\eg a logo or a product name), while other layers are live, encapsulating generative specifications that can be re-evaluated to produce new content. Authors can derive a layer from another one --- creating a dependency between them --- to refer to its content as part of the generative specification. These dependencies are intentionally constrained to single parent, so that the artifact’s behavior remains acyclic and easily interpretable.

\subsection{Spatial Composition}
\label{sec: Spatial Composition}
The canvas provides a visual canvas as workspace in which multimodal elements are represented as manipulable layers that can be positioned, resized, and reordered along the Z-order (\autoref{fig: interface}).  In the coffee ad example, the author composes the coffee shop logo, tagline, and an illustration together (\autoref{fig: interface}-C). When tagline and background are re‑evaluated, the logo and spatial composition remain fixed, preserving the overall look-and-feel of the artifact.

As generative exploration can become messy and divergent, the authoring tool provides two spaces for separating freeform ideation from composition (\autoref{fig: interface}-A). Users may freely create and modify layers across the infinite exploration canvas, but only layers dragged within the bounds of a dedicated output space become part of the resulting Live Artifact. The Layer stack only includes layers that are part of an output space (\autoref{fig: interface}-B), to limit visual complexity and to enable managing the Z-order depth and opacity of the artifact layers. This separation allows authors to treat the broader canvas as an experimentation space for iterating over generative specifications, for authoring components with different modalities, and for deciding which layers to keep live and which to incorporate as static media.

\subsection{Live Layers}
\label{sec: Live Layers}
In LiveCanvas, a static layer stores pixels, text, audio or video while a live layer stores a generative specification that can be re-evaluated (\textbf{C1}). In the coffee ad example, the shop logo is a static layer (\autoref{fig: interface}-B, \includegraphics[width=0.03\linewidth]{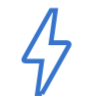}), while the tagline and illustration are live layers (\autoref{fig: interface}-B, \includegraphics[width=0.03\linewidth]{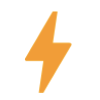}). The specification for the text is re-evaluated leading to a different drink idea given the location and weather at the time, which in turns triggers the re-evaluation of the derived image for this drink.

Authors can use live layers as one-shot output, turning it static when they reach a specific outcome (Literal, \includegraphics[width=0.03\linewidth]{figures/liveoff.png}); they can leave it live for regeneration (Live, \includegraphics[width=0.03\linewidth]{figures/liveon.png}), or  on automatic regeneration over time (Self-Regenerate, \includegraphics[width=0.03\linewidth]{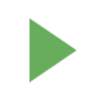}). By cycling through these states, authors can refine specifications and decide what remains stable and what changes, defining the exact bounds of the artifact identity and its non-deterministic content.

\subsection{Contextual Layers and External Data}
\label{sec: Contextual Layers}
Live Artifacts respond to context by integrating external data sources directly in the generative specification (\textbf{C2}). These specifications can reference location, or data sources returned from a live API or webpage. For example, in the coffee ad example, the tagline references a URL providing geographic localization and weather as part of its specification (\includegraphics[width=0.03\linewidth]{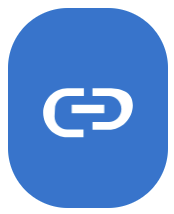} in \autoref{fig: interface}-D). During re-evaluation, the system fetches and parses this data and uses it as contextual grounding when generating the tagline and imagery.
Context is consumed as a lightweight inline reference in the specification that is re-evaluated over time or at each viewer consumption.

\subsection{Dependencies Between Layers}
\label{sec: Dependencies}
Live layers define what may vary; dependencies define how variation propagates (\textbf{C3}). 
Dependencies declare child-parent relationships describing how content flows between layers. 
Authors create a dependency by deriving one layer from another using a contextual menu (\autoref{fig: interface}-D), which injects a reference of the parent content (\includegraphics[width=0.03\linewidth]{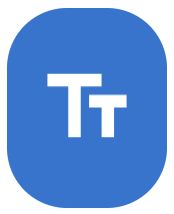} in \autoref{fig: interface}-D) directly into the child’s specification. When the parent re-evaluates, dependent layers re-evaluate as well, maintaining coherence across modalities. Dependencies are shown as connection lines on the canvas and in the layer stack.

For example, when the coffee ad tagline are re-evaluated based on live weather data, its content is propagated to the specification of the derived image layer. In this case, a new beverage type depending on current weather is populated and used to regenerate its illustration (\autoref{fig: example1}-a). This coordination comes from the dependency link and integration of the text layer reference in the child image specification, not from authored if-then rules.

To create a visually unified and consistently styled visual result even when combining diverse visual source material, the system creates a unifying merging layer from a selection of multiple layers. It takes the spatially arranged selected layers and compiles them into a unified visual composite layer (preserving their relative coordinates, scale, and transparency). This merged layer is recomputed after each regeneration of individual layers and does not destruct the underlying layer composition.

\subsection{LiveCanvas Constraints}
\label{sec: Media Abstraction}

We designed LiveCanvas around a set of constraints to preserve a media-first authoring model.

First, the system is organized around re-evaluation rather than execution. We do not support implementation of interaction logic, authored branching, or multi-step procedural workflows. Adding these constructs would require state management and debugging abstractions that depart from the simplicity of this interface targeted at non-programmers.
Thus, dependencies between layers are limited to ensure that the whole dependency graph remains acyclic. Each live layer is derived from a single parent.
When authors need coherence across multiple elements, they can create a merge layer to construct a single combined parent layer. Merge layers cannot be merged with other layers, however they can be used as a parent layer (used in the specification of a child layer). This decision is a tradeoff between flexibility (the ability to compose layers from other layers) and legibility (the number and complexity of dependency links on the screen), but not a limitation of the backend.

Second, timing is orchestrated rather than conditional. Authors do not script event listeners or encode fine-grained thresholds such as ``\textit{update only if temperature changes by 5\%.}'' Instead, layers re-evaluate through system-supported refresh behaviors, such as consumption-time refresh or simple intervals. This prioritizes simplicity and supports ambient liveness without requiring authors to reason about hidden execution states.

\subsection{System Architecture and Implementation}
\label{sec: Implementation}
The authoring environment is implemented as a web-based spatial canvas (TypeScript/ReactJS) backed by a Python/Flask service layer. To support the media-first abstractions described above without exposing visual programming wires, the system relies on specific backend data structures and a dynamic model routing engine.

\textbf{Data structures and artifact schema.}
While the frontend renders traditional spatial layers, the backend serializes the artifact as a Directed Acyclic Graph (DAG) of computational wrappers. Each compiled layer is stored as a schema tuple: (Spatial Bounds, Semantic Specification, Execution Mode, Dependency Links). The Execution Mode acts as a strict execution constraint at the system level: the backend is only permitted to request a new API generation if the layer's state evaluates to Live or Self-Regenerate.

\textbf{Resolving dependencies.}
The backend automatically serializes relationships between layers into programmatic dependencies:

\begin{itemize}
    \item Inline tokens: When a user visually derives a child from a parent, the backend injects the parent's generated content array into the child's generative schema as an inline token. When the temporal engine flags a parent state update, it traces this token to automatically trigger the child’s recalculation; 
    \item Merge node serialization: 
    This Merge operation calculates the relative X/Y coordinates, scale, and alpha transparency of selected layers, flattening them into a unified visual matrix. This matrix is serialized as a single, centralized parent node in the DAG. This allows the backend to provide holistic contextual grounding for downstream layers without ever needing to resolve multi-parent logic paths.
\end{itemize}

\textbf{Re-evaluation triggers.}
In conventional GenAI workflows, execution is a discrete, user-triggered event. Live Artifacts, however, demand continuous, context-driven evolution. 
LiveCanvas triggers reevaluation of the Live Artifact following the two conditions:

\begin{itemize}
    \item Manual refresh (each time the viewer consumes the artifact): Triggers a single traversal of the graph. The system identifies all ``Live'' layers, resolves their dependencies sequentially via their inline tokens, and updates the canvas simultaneously.
    \item Autonomous polling: For layers set to the ``Self-Regenerate'' mode, a temporal wrapper executes an interval-based loop (\eg each hour). The backend continually evaluates the layer's semantic prompt including polling external data APIs (\eg weather JSONs) and autonomously fires a topological traversal when the external data condition changes.
\end{itemize}

\textbf{Modality-agnostic layers}. 
An important architectural decision in \system~ is treating layers as modality-agnostic nodes within a unified dependency graph. Any text, image, audio, video, or sketch layer can function as parent or child layer. For instance, authors can specify a live text layer by referring to drawing on a sketch layer, or by using the content of an audio layer. 

Since layers are modality-agnostic, the backend acts as a specialized routing engine orchestrating all transformations across modalities required across the following models:

\begin{itemize}
    \item Text and Logic Processing: Semantic reasoning, textual generation, and API data parsing are handled by GPT-4o~\cite{openai_gpt4o};
    \item Auditory Synthesis: Text-to-speech and speech-to-text generation are routed through gpt-4o-mini-tts~\cite{openai_gpt4otts} and gpt-4o-mini-transcribe~\cite{openai_gpt4otrans};
    \item Imagery and Spatial Context: Base raster generation is powered by Stable Diffusion~\cite{rombach2022stablediffusion}, and merged‑layer image generation is supported by GPT‑Image~\cite{openai_gpt4o_image}. To handle the complex spatial overlaps required by our Dual‑Space architecture (\eg collages), transparent image requests are specifically routed to LayerDiffusion~\cite{li2023layerdiffusion} and ControlNet~\cite{zhang2023controlnet}, which provide latent‑space transparency control;
    \item Sketch and Conditional Anchors: Sketched layers are processed as conditional image-to-image inputs via ControlNet~\cite{zhang2023controlnet}, transforming manual strokes into structural anchors for generation;
    \item Video Execution: Text-to-video generation is processed via OpenAI Sora~\cite{sora}.
\end{itemize}

\section{User Study}

We conducted an exploratory qualitative study to understand how creators reason about Live Artifacts as an emerging genre of media, including the aspects they find compelling and the applications they imagine. Grounded in use of LiveCanvas, the study also reveals preliminary insights into an intermediary, media-first authoring mindset focused on specifying structure and behavior, rather than writing programs.

\subsection{Method}
We recruited six participants (P1–P6; 4f/2m) from diverse creative backgrounds, including design, animation, filmmaking, and creativity research. Participants were active multimedia creators, producing content at least weekly (one daily), with experience spanning graphic and UX design, presentations, video, interactive media, and product or wearable design. 
While all participants regularly used professional creative tools (\eg Photoshop, Figma, Illustrator, PowerPoint, Blender, Rhino, TouchDesigner), their familiarity with programming and code-based workflows varied. 
Participants P2, P4, and P6 had limited prior experience with programming, whereas P1, P3, and P5 had more substantial experience using code and computational tools as part of their creative practice. All participants routinely used generative AI systems such as ChatGPT, Midjourney, Stable Diffusion, Runway, and Copilot.

We employed a two-phase qualitative study. In Phase 1, participants took part in an onboarding session that introduced the concept of Live Artifacts and presented system-generated examples to establish a shared conceptual understanding of \media. Participants then discussed initial impressions, perceived possibilities, and how similar artifacts might be created using their existing workflows. 
In Phase 2, participants authored a simple Live Artifact with assistance from the experimenter at first; then created an artifact on their own based on one of their ideas. This session was followed by a semi-structured interview focusing on participants’ experiences and comparative reflections between authoring Live Artifacts and their existing practices, including both traditional media tools and AI-mediated workflows such as prompt-based generation or vibe coding. Each session lasted approximately 90 minutes; participants received \$50 compensation.

\begin{figure*}[!htb]
  \centering
    \centering
    \includegraphics[width=0.93\linewidth]{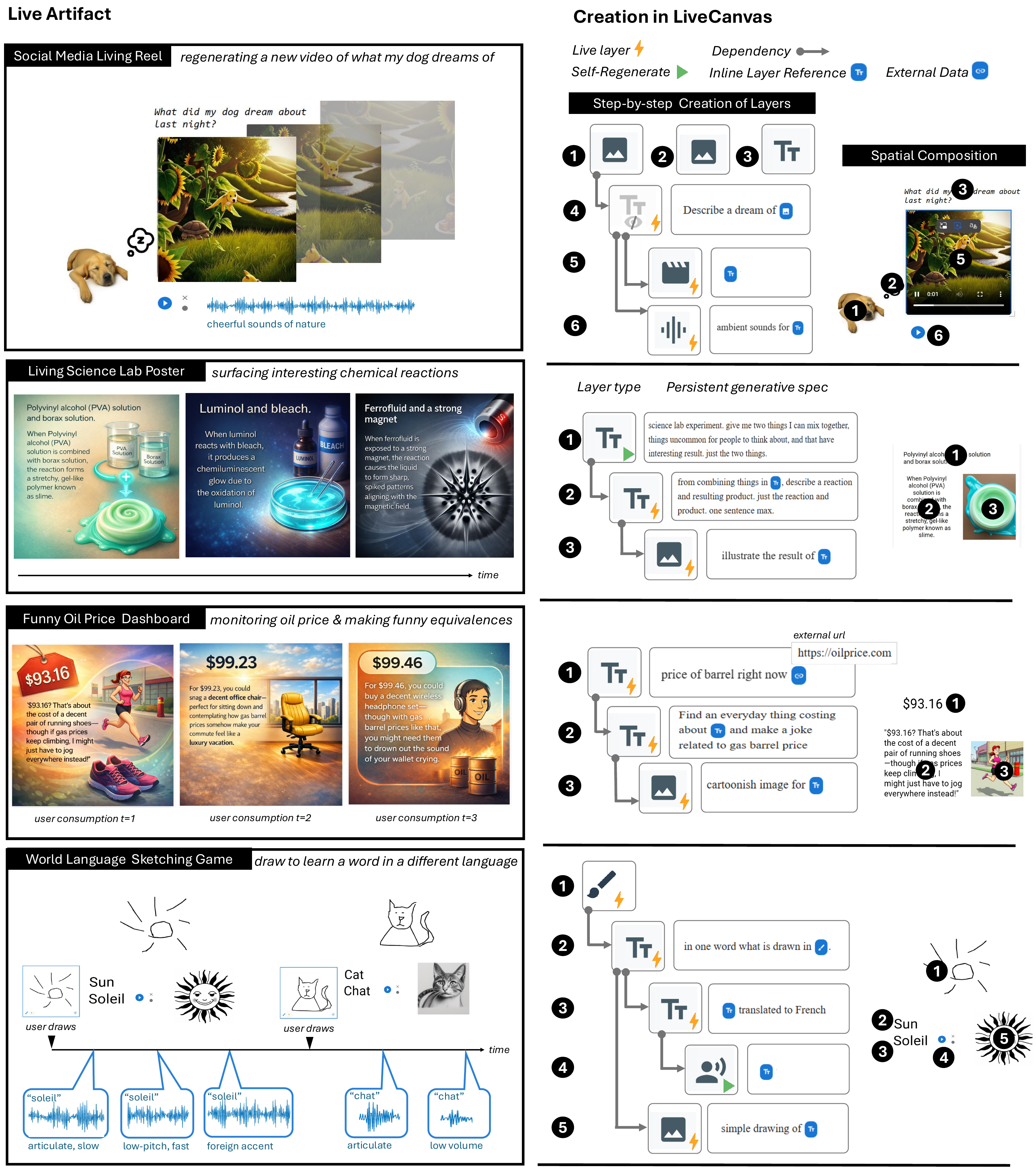}
     \vspace{-5px}
    \caption{Gallery of Live Artifacts created during the user study using \system. On the left is the Live Artifact; on the right its layer composition (layer type with generative specification, and spatial composition). The social media reel combines static and live layers, generating video and audio through an invisible intermediary live layer that produces a dream-like description based on the user’s uploaded pet image. The living science poster is self-reevaluating over time, leveraging the non-deterministic nature of models to surface different chemical reactions. The oil dashboard illustrates how contextual responsiveness can support monitoring external data as a funny custom live artifact. The sketch-to-translation artifact is triggered for each viewer’s sketch; the audio live layer re-evaluates itself over time leveraging the model’s non-determinism to render the word in varying voices and tones.}
\label{fig: gallery}
  \vspace{-10px}
\end{figure*}

\subsection{Findings}

Participants created a variety of Live Artifacts using LiveCanvas, spanning different scenarios and expressive goals; a subset of these artifacts is shown in the example gallery (\autoref{fig: gallery}). Reflecting on these scenarios, interviews highlighted which aspects participants found compelling and how authoring Live Artifacts reshaped their practices relative to their own practices (with both static media creation tools and programming).

\paragraph{\textbf{Experiencing “liveness”: novelty without losing identity}}
Participants consistently described Live Artifacts as capable of ongoing variation while preserving a stable, recognizable identity. Rather than framing regeneration as disruptive, they emphasized evolution within authored bounds, where composition, style, and structure remain legible across realizations. This was reflected in how participants imagined adaptation across contexts. P2 noted the potential for targeted advertising, suggesting that an artifact could adapt \textit{``into many versions for different audiences''} without manual duplication. In educational settings, P4 similarly described adapting content within a shared structure, observing that \textit{``combining user interests to illustrate examples is an excellent touch,''} and noted that in situated environments such as public installations or museums, \textit{``continuous contextual changes will bring real-time informational feedback to users, emphasizing their ‘presence’ in the space.''}

Several participants highlighted the experiential qualities of the dynamic media artifacts they created. P1 described how repeated encounters could yield moments of serendipity, noting that Live Artifacts offered \textit{``new inspirations that hit the user without them actively rolling the dice.''} Others emphasized the value of subtle, longitudinal change. P6 characterized this as a peripheral experience, explaining that \textit{``it’s not strictly purpose-driven, but occasionally you glance at it, and it has subtly changed.''} Across these accounts, participants interpreted Live Artifacts not as interchangeable outputs, but as enduring media objects that regenerate without losing identity. LiveCanvas supported this perception by localizing variation to specific layers while preserving overall spatial composition, enabling artifacts to feel both stable and continually renewed.

\paragraph{\textbf{Authoring ``behavior'': shaping change, not outputs}}
Participants described authoring Live Artifacts as a shift from producing static outputs or implementing programs toward specifying how an artifact should behave across repeated realizations. Rather than treating generation as a one-shot operation, authors defined constraints, relationships, and structural invariants that govern variation while preserving identity. As P3 explained, \textit{``the interactive paradigm has changed. It’s no longer linear thinking; it’s flattened. There is no strict sequential order; you just link their logical relationships.''} Authoring was thus framed not as assembling a result step by step, but as designing a persistent generative specification that shapes behavior over time.

This shift was also reflected in participants’ use of spatial layout and placement. Instead of positioning elements for a final composition, participants reasoned about how elements relate logically within the artifact. As P6 noted, \textit{``My attention shifted away from ‘where should I leave white space for text.’ Because the content is unfixed, your focus moves to how elements are logically associated.''} Layout, in this sense, became more a loose constraint rather than a meticulously optimized task given the more uncertain re-evaluation of live layers.

Compared to classic static media workflows, participants valued the cross-modality capabilities;  P5 explained: ``\textit{aligning different modalities is currently a massive barrier. You have to export from one platform, change formats, and import into another… the workload is huge}.'' However participants also acknowledged that relinquishing pixel-final determinism was a tradeoff. Because content may be re-evaluated across time or context, not every visual detail can be fixed in advance which could be problematic for high-stakes scenarios. As P3 noted, \textit{``the system applies a strong set of rules on top of the generation, ensuring it always adheres to a predefined architecture''}, signaling that while the artifact preserved a core architecture, it still had unpredictable outcomes. Participants thus viewed Live Artifacts as complementary to static media --- trading pixel-level finality for structural consistency.

When contrasted with programming or prompt-centric vibe coding, participants described a different boundary. Rather than repeatedly adapting layout and intent through trial and error, authors worked with visible constraints on the canvas. P4 described compensating for fragile prompt-based control by generating \textit{``5 to 10 images repeatedly until the layout accidentally matches the requirement.''} In contrast, P5 highlighted the immediacy of direct manipulation, noting that \textit{``the WYSIWYG aspect is incredibly fast. Instead of making a prototype, sending it for review, and coming back to tweak prompts, everyone can just look at the canvas and adjust the specific layer instantly.''} Participants understood the reduced expressive power compared to general‑purpose programming as a deliberate tradeoff, valuing the resulting legibility of generative specifications while acknowledging that building more complex logic would necessitate traditional programming.

These insights provide an early view into how creators reason about this genre of media, articulating perceived experiential qualities of Live Artifacts as well as emerging shifts in authoring practice at the boundary between static media and programming.

\section{Discussion and Future Work}

This research surfaces a set of design tradeoffs that emerge when generative logic is treated as a persistent property of media artifacts, rather than as an ephemeral rendering step or as full application logic. By examining these tradeoffs, we articulate the implications of persistence, constraint, and generative autonomy, and the tensions they introduce for future generative authoring systems.

\paragraph{\textbf{Specifications as a Media Property vs. One-Shot Generation}}

A central implication of this work concerns where generative logic lives. Most current generative workflows treat generation as an authoring operation that produces a static result, after which the logic that produced it is discarded.  Systems that retain generative logic typically do so in the form of software (\eg scripts, applications, or pipelines), foregrounding execution, state, and correctness while treating media properties such as visual appearance, layout, and composition as secondary outcomes of execution.

Live Artifacts occupy a different point in this space by embedding selected generative specifications directly into the artifact itself. Rather than collapsing generation into a one‑shot workflow, persistence becomes a property of the media. Our findings suggest that this persistence is valued when it supports varied, context‑aware viewing experiences while preserving a recognizable identity, and when it reduces the repetitive effort required to author variations with static media. In this sense, persisting generative specifications adds value both to how artifacts are experienced over time and to how they are authored and maintained. 

\paragraph{\textbf{Structural vs. Pixel‑precise Control}}
Once generative logic becomes a persistent property of media, a question is what form of control authors retain over change. Live Artifacts trade pixel‑precise control for structural consistency: authors relinquish exact control over final outputs, but gain control over composition, layout, and semantic relationships of regenerated outputs. Rather than fixing individual outcomes, authors define constraints that preserve identity, style, and architectural intent across re-evaluations, allowing variation to occur within recognizable bounds.
At the same time, this tradeoff implies clear boundaries on where Live Artifacts are appropriate. For scenarios that demand strict predictability, precise reproduction, or where authorial credibility hinges on fully specified outcomes such as legal, scientific, or high‑stakes commercial media, pixel-final determinism remains essential. In these contexts, nondeterministic regeneration may be undesirable regardless of structural constraints. This suggests that the challenge is not adopting nondeterminism per se, but designing interfaces that make constraints visible and understandable as part of the artifact.

\paragraph{\textbf{Expressiveness vs. Legibility in Authoring}}

Another important tradeoff concerns balancing expressive power with how easily authors can grasp and use the system. LiveCanvas deliberately constraints dependencies between layers to limit complexity and keep the layer stack legible. This choice surfaces a recurring HCI tradeoff: increased expressiveness often comes at the cost of legibility and direct manipulation~\cite{Satyanarayan2019a}.

Participants contrasted LiveCanvas with both prompt-centric generation and code-based approaches. While programming offers greater expressive power, it shifts attention toward implementation concerns such as execution order, debugging, and failure cases. Prompt-based approaches, while accessible, were described as fragile and difficult to coordinate spatially or across modalities. In contrast, LiveCanvas emphasized legible relationships (visible layers and bounded dependencies) that allowed participants to reason locally about how changes would propagate.

The implication is less about replacing programming than about the value of an intermediate representational form. By retaining persistence and propagation while excluding full control flow and state, Live Artifacts offer a space where authors can specify durable generative specifications without adopting an engineering mindset. The value of this middle ground lies in supporting reasoning about change and continuity using media-native constructs.

\paragraph{\textbf{Augmenting Media vs. Simplifying Programming}}

A key design tension raised by Live Artifacts concerns whether the goal is to augment media with generative capabilities, or to simplify programming by making it more visual and accessible. Although these directions may appear similar on the surface, they imply different priorities and boundaries. Live Artifacts are intentionally stateless in the software sense: content changes through re‑evaluation rather than through accumulating internal state. This choice preserves a media‑first mental model, where authors reason about composition, structure, and relationships rather than execution and process. At the same time, it constrains both temporal continuity and the kinds of user interaction that can be supported without introducing programmatic control.

Participants quickly identified scenarios where some form of memory, accumulation, or responsiveness to viewers would be desirable, such as artifacts that evolve gradually over time or acknowledge repeated user input. Supporting these behaviors, however, would require moving beyond simple re‑evaluation toward mechanisms for retaining state, handling events, and branching behavior across interactions. At that point, shifting toward a programming model may be both appropriate and necessary but it also entails a qualitative change in how authors reason about the artifact. This would reintroduce many of the programming constructs that Live Artifacts are designed to avoid, along with the loss of directness and media‑first legibility that motivates the approach.

This tension signals a design boundary. Live Artifacts are not an attempt to make programming easier through visual abstraction; rather, they explore how far media itself can be augmented with persistent generative behavior before a transition to programming becomes the more suitable choice. The question is not whether programming constructs can be added, but whether temporal continuity and limited forms of user interaction can be supported through constrained, media‑native mechanisms without crossing into programming.  Where this boundary lies, and what is gained or lost in crossing it, remains an open question.

\paragraph{\textbf{Limitations.}}
In practice, making Live Artifacts feel truly “live” is limited by the latency of cloud-based generation models. This is rarely an issue for ambient or background updates, but it becomes a bottleneck for real-time interactions. Hybrid approaches — such as using local models for fast previews and cloud models for final output— could help reduce this friction. 

While managing many live layers may become challenging as artifacts grow in complexity, our study and gallery already suggest that artifacts with relatively few layers can be expressive and compelling. Scaling to larger artifacts would likely benefit from established structuring techniques — such as layer grouping and hierarchical organization~\cite{xia2017} — to preserve legibility of dependencies without overwhelming authors.

Our study offers early insights into how people reason about Live Artifacts, as well as preliminary feedback on authoring them as Live Layers in LiveCanvas. As a creative medium, Live Artifacts need to be placed in people’s hands over longer periods and in real-world contexts to understand where they provide sustained value and what use cases emerge organically in practice. Deeper investigation is also needed to examine the design tradeoffs we made in LiveCanvas (\eg single‑parent dependencies, simplified temporal reevaluation) to find the right balance between creative freedom, legibility, and media‑centric reasoning. Future work could probe at the broader design space of authoring paradigms for outputs bridging static media and full‑fledged software.

\vspace{-.2cm}
\section{Conclusion}

This paper introduces \textit{Live Artifacts}, a class of generative media in which parts of the generative specification persist as an ongoing property of the artifact, enabling regeneration that can respond to time, context, or live data while preserving recognizable structure and cross‑modal coherence. 
By extending the familiar abstraction of visual layers to be live and linkable, \system~ demonstrates how these properties can be authored within a media‑first workflow grounded in spatial composition and direct manipulation. Insights from an exploratory study offer preliminary perspective on the potential and range of examples for Live Artifacts, along with the authoring tradeoffs of augmenting media with generative persistence without shifting into programming abstractions.  This research outlines a design space for persistent generative media that complements both traditional media authoring and software development, and point to opportunities for rethinking how generative behavior might be authored through media‑native mechanisms, rather than through programming‑ or chat‑centric abstractions.

\bibliographystyle{ACM-Reference-Format}

\bibliography{main}

\clearpage

\end{document}